\documentclass[11pt, a4paper]{article}
\usepackage[T1]{fontenc}
\usepackage{a4wide, url}
\usepackage[english]{babel}
\usepackage{graphicx}
\usepackage{array}
\usepackage{natbib}
\usepackage[textfont=sc,labelfont=bf]{caption}
\usepackage{setspace}
\usepackage{multirow}
\usepackage{amsthm}
\usepackage{amsmath}
\usepackage{amsfonts}
\usepackage{amssymb}
\usepackage{bigints}
\usepackage{pgf} 
\usepackage{bm}
\usepackage{paralist}
\usepackage{hyperref}
\usepackage{pdflscape}
\usepackage{xfrac}
\usepackage[linesnumbered,ruled,vlined]{algorithm2e}
\SetKwInput{KwInput}{Input}                
\SetKwInput{KwOutput}{Output}              
\usepackage{enumitem,kantlipsum}
\usepackage{xr}
\usepackage{yfonts}

\usepackage{xcolor}
\hypersetup{
    colorlinks,
    linkcolor={red!50!black},
    citecolor={blue!50!black},
    urlcolor={blue!80!black}
}

\usepackage{comment}
\excludecomment{ext}\excludecomment{comment}

\usepackage[top=1in,bottom=1in,left=.75in,right=.75in]{geometry}

\newcolumntype{L}[1]{>{\raggedright\let\newline\\\arraybackslash\hspace{0pt}}m{#1}}
\newcolumntype{C}[1]{>{\centering\let\newline\\\arraybackslash\hspace{0pt}}m{#1}}
\newcolumntype{R}[1]{>{\raggedleft\let\newline\\\arraybackslash\hspace{0pt}}m{#1}}

\DeclareMathAlphabet\mathbfcal{OMS}{cmsy}{b}{n}

\newcommand{\beq}{\begin{equation}}
\newcommand{\eeq}{\end{equation}}
\newcommand{\bea}{\begin{eqnarray}}
\newcommand{\eea}{\end{eqnarray}}
\newcommand{\ba}{\begin{array}}
\newcommand{\ea}{\end{array}}
\newcommand{\bit}{\begin{itemize}}
\newcommand{\eit}{\end{itemize}}
\newcommand{\ben}{\begin{enumerate}} 
\newcommand{\een}{\end{enumerate}}
\newcommand{\bpm}{\begin{pmatrix}}
\newcommand{\epm}{\end{pmatrix}}
\newcommand{\bbm}{\begin{bmatrix}}
\newcommand{\ebm}{\end{bmatrix}}

\newcommand\norm[1]{\left\lVert#1\right\rVert} 
\usepackage{booktabs} 
\usepackage{mathpazo} 
\usepackage{subcaption} 
\usepackage{fancybox,framed} 
\usepackage{float} 
\usepackage{relsize} 
\usepackage{mathtools} 
\usepackage{subcaption}

\title{\textsc{\large 	
{Predicting Energy Demand
with Tensor Factor Models}}}
\date{}

\makeatletter
\renewcommand\@makefnmark{}  
\makeatother

\begin{document}
\maketitle

\begin{center}\vspace{-1.5cm}
\begin{tabular}{cp{1cm}cp{1cm}c}
 Mattia Banin && Matteo Barigozzi & &Luca Trapin \\[.1cm] 
\footnotesize Anima Sgr && \footnotesize Universit\`a di Bologna && \footnotesize Universit\`a di Bologna
\\[.5cm]
\end{tabular}

\small This version: March, 2026

\footnotetext{Disclaimer: the views expressed in this paper are those of the authors and do not necessarily reflect the views and policies of Anima Sgr. }
\footnotetext{
	Acknowledgments: LT acknowledges financial support under the National Recovery and Resilience Plan (NRRP), Mission 4, Component 2, Investment 1.1, Call for tender No. 1409 published on 14.9.2022 by the Italian Ministry of University and Research (MUR), funded by the European Union --NextGenerationEU-- Project Title ``Measuring, managing and hedging indirect climate-transition risks'' --CUP P20228CHNL-- Grant Assignment Decree No. 1376 adopted on 01/09/2023 by the Italian Ministry of Ministry of	University and Research (MUR).
}
\end{center}

\begin{abstract}
{Hourly consumption from multiple providers displays pronounced intra-day, intra-week, and annual seasonalities, as well as strong cross-sectional correlations. We introduce a novel approach for forecasting high-dimensional U.S. electricity demand data by accounting for multiple seasonal patterns via tensor factor models. 
To this end, we restructure the hourly electricity demand data into a sequence of weekly tensors. Each weekly tensor is a three-mode array whose dimensions correspond to the hours of the day, the days of the week, and the number of providers. This multi-dimensional representation enables a factor decomposition that distinguishes among the various seasonal patterns along each mode: factor loadings over the hour dimension highlight intra-day cycles, factor loadings over the day dimension capture differences across weekdays and weekends, and factor loadings over the provider dimension reveal commonalities and shared dynamics among the different entities. We rigorously compare the predictive performance of our tensor factor model against several benchmarks, including traditional vector factor models and cutting-edge functional time series methods. The results consistently demonstrate that the tensor-based approach delivers superior forecasting accuracy at different horizons and provides interpretable factors that align with domain knowledge. Beyond its empirical advantages, our framework offers a systematic way to gain insight into the underlying processes that shape electricity demand patterns. In doing so, it paves the way for more nuanced, data-driven decision-making and can be adapted to address similar challenges in other high-dimensional time series applications.
}
\vspace{0.5cm}


\end{abstract}

\renewcommand{\thefootnote}{$\ast$} 
\thispagestyle{empty}


\renewcommand{\thefootnote}{\arabic{footnote}}


\newpage
\section{Introduction}

{The increasing availability of high-frequency, high-dimensional time series data has opened new opportunities for analysts, forecasters, and decision-makers across numerous domains. These modern datasets often contain dozens or even hundreds of related series, each measured at fine temporal resolutions—such as hourly or sub-hourly frequencies—and governed by intricate underlying patterns. In many settings, the observed series exhibit multiple, interacting forms of seasonality, including intra-day, intra-week, and annual cycles. Such complexities arise across a wide range of applications, from modeling call center arrivals, hospital admissions, or web traffic, to understanding retail transactions and supply-chain dynamics \citep[see, e.g.,][]{gould2008forecasting, taylor2010exponentially, de2011forecasting}.}

{A particularly pertinent example is electricity demand forecasting. The widespread adoption of smart meters has enabled utilities to collect highly granular consumption data at increasingly large scales, capturing fine-grained usage patterns, at an hourly or even sub-hourly level, for entire regions, multiple providers, or even individual households. As a result, modern electricity datasets are rich both cross-sectionally and temporally. This setting naturally involves multiple seasonalities: intra-day cycles reflect human activity rhythms, intra-week variations distinguish between weekdays and weekends, and annual cycles coincide with long-term climatic and economic factors. Moreover, these multi-level seasonalities interact with cross-sectional variation as utilities, balancing areas, and regions exhibit correlated but distinct demand patterns. Accurate forecasting of this high-dimensional, multi-seasonal structure is essential for efficient and sustainable grid management, as it underpins resource allocation, demand response strategies, and long-term infrastructure planning \citep{harvey1993forecasting, cottet2003bayesian, smith2010modeling, cho2013modeling, taieb2021hierarchical}.
}

{Factor models have proven to be powerful tools for summarizing the common dynamics among large collections of time series. They reduce dimensionality by identifying latent factors that drive co-movements in the data. As datasets have grown in complexity, traditional vector factor models (VFMs) have evolved into more flexible formulations that accommodate multiple dimensions. Matrix factor models (MFMs) extend VFMs by using matrix representations to exploit additional structure, while tensor factor models (TFMs) generalize this approach even further by representing data as multi-way arrays (tensors). In these richer frameworks, each dimension (or mode) of the tensor can capture a distinct source of variation, such as time, location, product categories, or other meaningful attributes. This progress is well documented in the econometrics and statistics literature, see, e.g., \cite{Wang2019, Chen2023, Yu2022} for MFMs, and \cite{Chen2022tensor, Chen2024, Zhang2024, Barigozzi2023b, Barigozzi2023a} for TFMs.}

{Building on these advancements, this work leverages TFMs to tackle the intricate problem of modeling multiple seasonalities in a high-dimensional setting. In the context of hourly electricity demand, we propose organizing the data into a weekly tensor, with one mode representing the hours of the day, another indexing the days of the week, and a third capturing the cross-section of providers or regions. Such a representation naturally encodes the multi-level seasonal structure, allowing a tensor factor decomposition to isolate and interpret different patterns: hourly loadings reveal intra-day cycles, daily loadings uncover intra-week fluctuations, and cross-sectional loadings highlight similarities and differences across providers. Using hourly U.S. electricity demand data as a case study, we show how our approach outperforms conventional methods in terms of forecasting accuracy while yielding interpretable factors and factor loadings that align with well-known domain patterns. Although we focus on electricity demand as a motivating example, the proposed methodology applies more broadly to a range of high-frequency, high-dimensional settings where multi-layered seasonal structures are present.} 

{The remainder of the paper is organized as follows. In Section~\ref{sec:model}, we present the tensor factor model and discuss its estimation procedure. Section~\ref{sec:data} describes the U.S. electricity demand dataset and highlights its main empirical characteristics. In Section~\ref{sec:empirical}, we apply the proposed methodology and compare its performance against competing benchmarks. }

\section{Electricity demand: the PJM dataset}
\label{sec:data}
\begin{figure}[htpb]
	\centering  
	\includegraphics[scale=0.90]{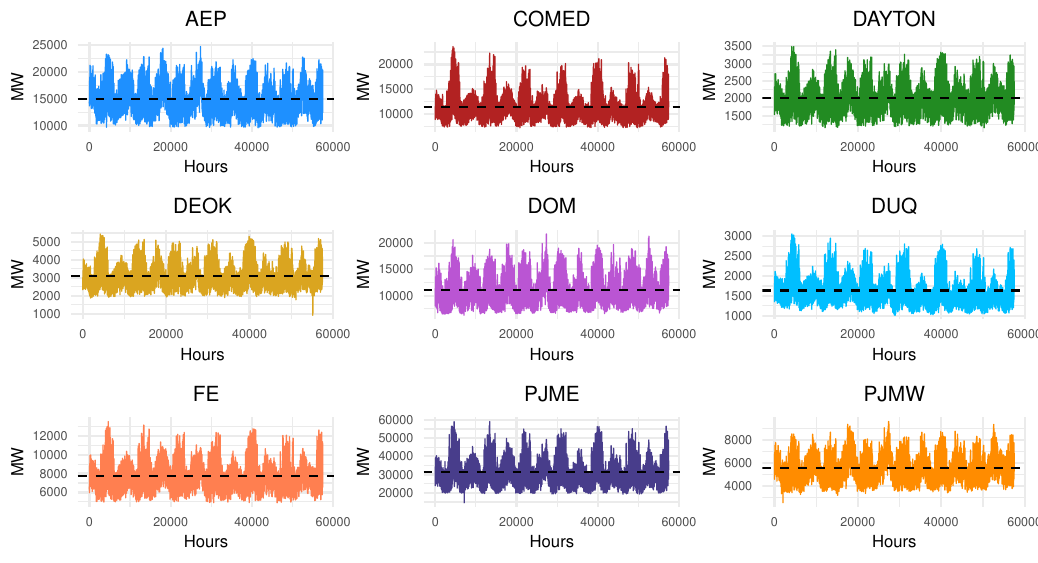} 
	\caption{Time series of hourly energy demand in Mega Watts} 
	\label{fig:time_series_dset_B} 
\end{figure}

\begin{figure}[htpb]
	\centering
	\includegraphics[width=0.8\textwidth]{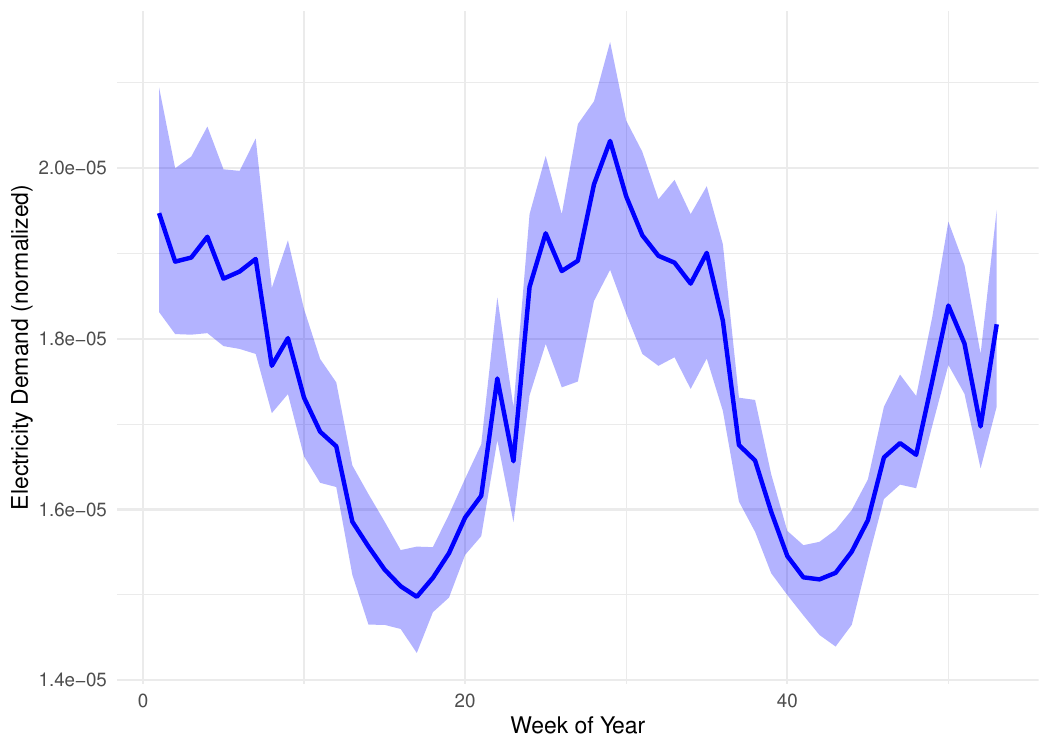}
	\caption{Weakly average (normalized) electricity demand over the year. Shaded areas and solid lines display the range and the average across companies, respectively.}
	\label{fig:IntraYearSeasonality}
\end{figure}

The data used in this work are produced by \textit{PJM Interconnection LLC}. PJM is a Regional Transmission Organization (RTO) coordinating energy distribution across 13 U.S. states and the District of Columbia.\footnote{More information available at \href{https://pjm.com/}{https://pjm.com/}.} For geographical reference, the 13 States are all located between the eastern part of the Midwest, the northern part of the Southeast, and the southern part of the Northeast.\footnote{The States totally or partially covered by PJM are Delaware, Illinois, Indiana, Kentucky, Maryland, Michigan, New Jersey, North Carolina, Ohio, Pennsylvania, Tennessee, Virginia, West Virginia, plus the District of Columbia.} The raw datasets used in this work are freely available at \href{https://kaggle.com/datasets/robikscube/hourly-energy-consumption/data}{this link}.\footnote{\href{https://kaggle.com/datasets/robikscube/hourly-energy-consumption/data.}{https://kaggle.com/datasets/robikscube/hourly-energy-consumption/data}} In total, data are available for 11 different electricity providers or groups of providers.\footnote{The geographical location of the providers is available under the \textit{Transmission Zones} button at \href{https://www.pjm.com/library/maps}{this link} and are labeled in the following manner: American Electric Power (AEP), Commonwealth Edison (ComEd),  Dayton Power and Light Company (DAYTON), Duke Energy Ohio/Kentucky (DEOK), Dominion Virginia Power (DOM), Duquesne Light Co. (DUQ), East Kentucky Power Cooperative (EKPC), FirstEnergy (FE), Northern Illinois Hub (NI), PJM East Region (PJME), PJM East Region (PJMW). In the latter two cases, PJME and PJMW are not unique providers but aggregates.} Hourly data between 2002 and 2018 in Megawatts (MW) are included. As most of the series does not record measurements starting from 2002, we only retain information on nine proiders (AEP, ComEd, DAYTON, DEOK, DOM, DUQ, FE, PJME, PJMW) from the beginning of January 2012 to the end of July 2018, collecting 57,456 hourly observations for each provider. \autoref{fig:time_series_dset_B}  displays the time series of the different providers. Despite data points being very dense, we can highlight a few interesting aspects that are relevant for model definition. Firstly, levels differ greatly between datasets. This is consistent with the geographical area of energy supply for the different providers: some companies operate in much larger areas than others. Secondly, the behavior of the series is strongly seasonal, and the absence of long-run trends is well-explained by the relatively short time period covered by the series (about 7 years). Some peaks are evident. Peaks are similar in magnitude and their distance from the others is regular, within the same time series. Across time series, the peaks are typically aligned over time. Interestingly, time series such as AEP, DEOK, and DOM present extremely regular over-the-mean peaks (\textit{regular-peaked series}). On the contrary, other time series such as COMED and DUQ present alternated over-the-mean peaks with different magnitude levels (\textit{alternate-peaked series}). All 9 time series can be easily categorized into one of these two categories. A closest qualitative analysis of the time series shows that all peaks (both for regular-peaked series and alternate-peak series) are semiannual. This is consistent with the typical behavior of electricity consumption. Indeed, semestral peaks correspond to the hottest weeks in summer, and the coldest weeks in winter. Typically, for the alternated peak series, the highest peaks are the ones in summer. This is probably due to the wide adoption of air conditioning in the U.S.A. for cooling offices, houses, and factories. On the other hand, heating systems do not only consume electricity but also natural gas. In general, even in regular-peaked time series summer peaks are slightly higher than winter peaks. This is further confirmed by the seasonal plot displayed in Figure \ref{fig:IntraYearSeasonality} showing the weakly average electricity demand for the companies, normalized by total company demand.

\begin{figure}[htpb]
	\centering
	\includegraphics[width=0.8\textwidth]{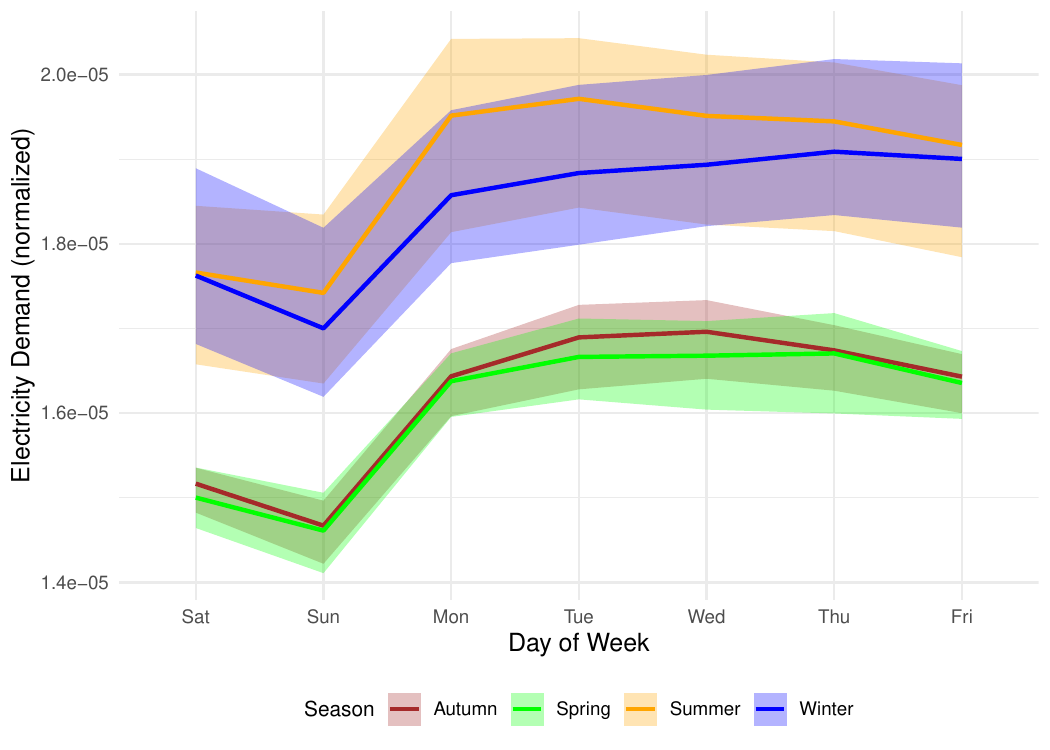}
	\caption{Average (normalized) electricity demand for every day of the week in each season. Shaded areas and solid lines display the range and the average across companies, respectively.}
	\label{fig:IntraWeekSeasonality}
\end{figure}

\begin{figure}[htpb]
	\centering
	\includegraphics[width=0.8\textwidth]{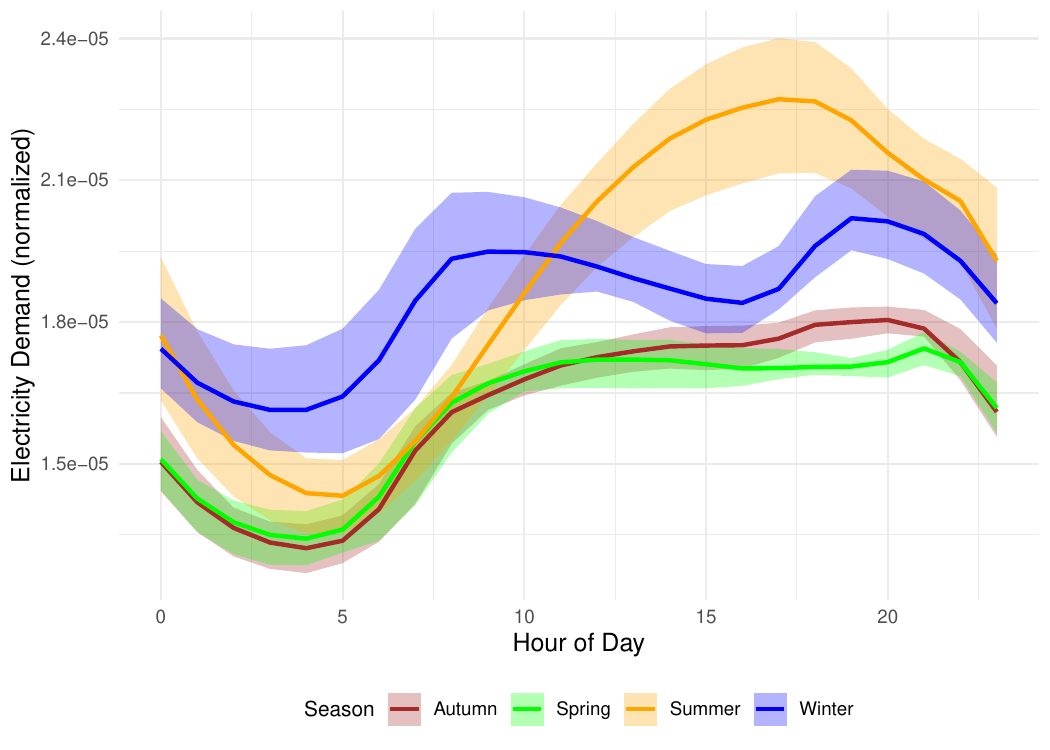}
	\caption{Average (normalized) electricity demand for every hour of the day in each season. Shaded areas and solid lines display the range and the average across companies, respectively.}
	\label{fig:IntraDaySeasonality}
\end{figure}

To investigate the multi-seasonal nature of electricity demand in greater detail, we also examine its intra-week and intra-day components. Figure~\ref{fig:IntraWeekSeasonality} shows the average (normalized) electricity demand by day of the week, grouped by the four seasons of the year. This approach allows us to assess the stability of the weekly demand pattern over the course of the year. The plot indicates that all providers exhibit very similar behavior: electricity demand is consistently higher on weekdays than on weekends, and this pattern remains broadly stable across all four seasons. While slight variations may occur, the overall day-of-week cycle does not appear to shift substantially between winter, spring, summer, and autumn. This finding underscores the robustness of the intra-week pattern, regardless of seasonal differences in temperature or other factors influencing electricity usage. Figure~\ref{fig:IntraDaySeasonality} illustrates the average (normalized) electricity demand of the companies for each hour of the day, again stratified by season. A pronounced intra-day pattern emerges, yet its shape and intensity vary across the year. In spring and autumn, demand typically rises late in the morning and remains elevated through the early afternoon. By contrast, summer consumption is generally higher overall, and demand continues to increase into the mid-afternoon (peaking around 4~p.m.) before gradually declining through the early morning hours. Winter usage is again higher than in spring or autumn but follows a more complex hourly profile compared to summer. Specifically, two peaks appear, one around 9a.m. and another around 8p.m., with a local minimum in the late afternoon (around 4~p.m.). This pattern likely reflects both increased heating needs and variation in daytime activities across different seasons.

Finally, to provide further information on the actual distributional characteristics of the time series, \autoref{tab:stat_summary} reports mean, median, standard deviation, skewness, and kurtosis for all of them. As expected, means and standard deviation differ considerably across companies. Moreover, all distributions are asymmetric and positively skewed, with sample kurtosis higher than three.

\begin{table}[ht]
\centering
\small
\begin{tabular}{lcccccc}
\toprule
\textbf{} & \textbf{Mean} & \textbf{Median} & \textbf{Standard Deviation} & \textbf{Skewness} & \textbf{Kurtosis} \\
\midrule
AEP    & 14998.6 & 14749  & 2501.355 & 0.428  & 2.806 \\
COMED  & 11383.48 & 11114 & 2278.45  & 1.131  & 5.038 \\
DAYTON & 2002.139 & 1973  & 378.478  & 0.518  & 3.142 \\
DEOK   & 3104.468 & 3012  & 600.309  & 0.680  & 3.365 \\
DOM    & 11049.34 & 10587 & 2433.582 & 0.733  & 3.263 \\
DUQ    & 1637.387 & 1597  & 303.561  & 0.857  & 3.961 \\
FE     & 7782.04 & 7693  & 1314.598 & 0.642  & 3.523 \\
PJME   & 31409.27 & 30479 & 6380.74  & 0.769  & 3.672 \\
PJMW   & 5575.884 & 5458  & 1009.528 & 0.455  & 2.899 \\
\bottomrule
\end{tabular}
\caption{descriptive statistics of time series data, dataset B}
\label{tab:stat_summary}
\end{table}

\section{Modeling High-Dimensional Time Series with Multiple Seasonalities}
\label{sec:model}
In this section, we introduce a unified factor-modeling framework capable of capturing multi-layered seasonal dynamics in high-dimensional time series. Subsection~\ref{sec:general_framework} describes the general approach, illustrating how multiple seasonal frequencies can be accommodated within a hierarchical factor model that simultaneously incorporates cross-sectional and seasonality-specific factors. Subsection~\ref{sec:hourly_model} then applies this framework to the problem of forecasting hourly electricity demand, showing how daily and weekly seasonalities, along with the cross-sectional dimension, can be naturally embedded in the model to provide a flexible and interpretable tool for analyzing complex energy consumption patterns.

\subsection{A General Modeling Framework}
\label{sec:general_framework}
Let $\{y_{i\tau}\}$ be a panel dataset consisting of $N$ time series observed over $\mathcal{T}$ time points. Here, $i = 1,\dots,N$ indexes the cross-sectional unit (e.g., an individual, region, or provider), while $\tau = 1,\dots,\mathcal{T}$ indexes time at the finest frequency. In many empirical applications, each time series $y_{i\tau}$ may exhibit multiple \emph{integer-valued} (i.e., calendar) seasonalities. For example, there might be an hourly pattern repeating every 60 minutes, a daily pattern repeated every 24 hours, a weekly pattern repeated every 7 days.

To accommodate these multi-layered \emph{calendar} seasonalities, we note that $\mathcal{T}$ can often be expressed as a product of distinct seasonal periods. More precisely, assume we have $M$  seasonal factors, each with an integer period $S_j$, where $j=1,\dots,M$. For example,  $S_1 = 7$ days in a week, $S_2 = 24$ hours in a day, and $S_3 = 60$ minutes in an hour, and so on. Under this framework, each time index $\tau$ can be mapped onto a collection of seasonal indices $(s_1,\dots,s_M)$ together with a lower-frequency index $t$:
\[
\tau \;\longmapsto\; (s_1,\dots,s_M,\, t),
\]
where $s_j = 1,\dots,S_j$ for each $j$, and $t=1,\dots,T$ with $T =\mathcal{T}/S$ and $S=\prod_{j=1}^{M}S_j$. Hence, rather than working directly with $\{y_{i\tau}\}$, one can re-index the data in a multi-way structure $\{y_{i,s_1,\dots,s_M,t}\}$. This re-indexing clarifies the position of each observation within multiple calendar-based cycles and partitions the total sample into $T$ complete arrays of these cycles.

To capture interacting seasonal patterns in a high-dimensional context, we propose a multi-level factor structure incorporating both cross-sectional and seasonality-specific factors. Let
$y_{i,\mathbf{s},t} = y_{i,s_1,\dots,s_M,t}$, and define a multi-level factor model of the form:
\begin{equation}
\begin{array}{rll}
	y_{i,\mathbf{s},t}  &=& \mu_{i,\mathbf{s}} + \sigma_{i,\mathbf{s}}  \epsilon_{i,\mathbf{s},t}, \\
	\epsilon_{i,\mathbf{s},t} &=& \sum\limits_{r=1}^R {\lambda}_{i,r} {f}^{(M)}_{r,\mathbf{s},t} + \nu_{i,\mathbf{s},t}, \\
	f^{(M)}_{r,\mathbf{s},t} &=& \sum\limits_{k_M=1}^{K_M} {\beta}^{(M)}_{s_M,k_M}  {f}^{(M-1)}_{r,s_1,\dots,k_{M},t} + \eta^{(M)}_{r,\mathbf{s},t},  \\
	{f}^{(M-1)}_{r,s_1,\dots,s_{M-1},k_M,t} &=& \sum\limits_{k_{M-1}=1}^{K_{M-1}} {\beta}^{(M-1)}_{s_{M-1},k_{M-1}} f^{(M-2)}_{r,s_1,\dots,k_{M-1},k_M,t} + \eta^{(M-1)}_{r,s_1,\dots,s_{M-1},k_M,t}, \\
	&\vdots& \\
	f^{(1)}_{r,s_1,k_2\dots,k_M,t}  &=& \sum\limits_{k_1=1}^{K_1}{\beta}^{(1)}_{s_1,k_1} {f}_{r,k_1,\dots,k_M,t} + \eta^{(1)}_{r,s_1,k_2\dots,k_M,t}.
\end{array}
\label{eq:gen_frame}
\end{equation}
where $\mu_{i,\mathbf{s}}$ is an intercept term capturing the individual-seasonal overall level, $\sigma_{i,\mathbf{s}}$ is parameter capturing the individual-seasonal scale, $\mathbf{f}^{(M)}_{\mathbf{s},t}$ is a $R$-dimensional vector of seasonal latent factors and $\boldsymbol{\Lambda}=[\lambda_{i,r}]$ is a $N\times R$ matrix of factor loadings characterizing the cross-sectional dependence among individuals, $\{{f}_{r,k_1,\dots,k_M,t}\}_t$ is a time series characterizing the temporal evolution of the latent factors, with $r=1,\dots,R$ and $k_j=1,\dots,K_j$, $\boldsymbol{B}^{(j)}=[\beta^{(j)}_{s_j,k_j}]$ is a $S_j\times K_j$ matrix of factor loadings describing the seasonal pattern of the $j$-th layer of seasonality. Multiple factors for each $j$-th layer might be necessary to characterize temporal variation in the $j$-th seasonal pattern. Finally, $\{\epsilon_{i,\mathbf{s},t}\}$ and $\{\eta^{(j)}_{k,\dots,t}\}$ are time series of idiosynchratic shocks for each individual and each layer of the seasonality, respectively. By arranging the data in this multi-index format and specifying distinct factor structures for each seasonal layer, we can flexibly account for multiple calendar-based cycles. The exact dimensionalities $k_j$, assumptions on the error terms, and identification constraints may vary depending on the practical application and estimation methods.

\paragraph{Remark.} Note that the multi-level factor model in \eqref{eq:gen_frame} can be written in the following compact form
\begin{equation}
\begin{array}{rll}
	y_{i,\mathbf{s},t}  &=& \mu_{i,\mathbf{s}} + \sigma_{i,\mathbf{s}}  \epsilon_{i,\mathbf{s},t}, \\
	\epsilon_{i,\mathbf{s},t}  &=&  \sum\limits_{r=1}^R {\lambda}_{i,k} \prod\limits_{j=1}^M \sum\limits_{k_j=1}^{K_j}{\beta}^{(j)}_{s_j,k_j} {f}_{r,\dots,k_j,\dots,t} + \varepsilon_{i,\mathbf{s},t}, \\
	\varepsilon_{i,\mathbf{s},t} &=& \sum\limits_{r=1}^R {\lambda}_{i,k} \left( \prod\limits_{j=1}^M \sum\limits_{k_j=1}^{K_j}{\beta}^{(j)}_{s_j,k_j} \eta^{(1)}_{r,s_1,k_2\dots,k_M,t} + \dots  + \sum\limits_{k_M=1}^{K_M}{\beta}^{(K_M)}_{s_M,k_M}  \eta^{(M-1)}_{r,s_1,\dots,s_{M-1},k_M,t} + \eta^{(M)}_{r,\mathbf{s},t} \right) + \nu_{i,\mathbf{s},t}.
\end{array}
\label{eq:gen_frame_comp}
\end{equation}
For example, consider a time series $y_{\tau}$ displaying two integer-valued seasonalities and assume a single factor driving both the seasonal temporal dynamics and the cross-sectional dependence, we can write
\[
	\begin{array}{rll}
		y_{i,s_1,s_2,t}  &=& \mu_{i,s_1,s_2} + \sigma_{i,s_1,s_2} \epsilon_{i,s_1,s_2,t} \\
		\epsilon_{i,s_1,s_2,t}  &=& {\lambda}_{i} {f}^{(2)}_{s_1,s_2,t} + \nu_{i,s_1,s_2,t}, \\
		f^{(2)}_{s_1,s_2,t} &=&  {\beta}^{(2)}_{s_2}  {f}^{(1)}_{s_1,t} + \eta^{(2)}_{s_1,s_2,t},  \\
		f^{(1)}_{s_1,t}  &=& {\beta}^{(1)}_{s_1} {f}_{t} + \eta^{(1)}_{s_1,t},
	\end{array}
\]
or, stated otherwise in compact form,
\[
	\begin{array}{rll}
		y_{i,s_1,s_2,t}  &=& \mu_{i,s_1,s_2} + \sigma_{i,s_1,s_2} \epsilon_{i,s_1,s_2,t} \\
		\epsilon_{i,s_1,s_2,t}  &=& {\lambda}_{i} {\beta}^{(2)}_{s_2}  {\beta}^{(1)}_{s_1} {f}_{t} + \varepsilon_{i,s_1,s_2,t}, \\
		\varepsilon_{i,s_1,s_2,t}  &=& {\lambda}_{i}  \left( {\beta}^{(2)}_{s_2}  \eta^{(1)}_{s_1,t} + \eta^{(2)}_{s_1,s_2,t} \right) + \nu_{i,s_1,s_2,t}.  \\
	\end{array}
\]

\subsection{A Model for Hourly Electricity Demand from Multiple Providers}
\label{sec:hourly_model}
We now specialize the general framework of Section~\ref{sec:general_framework} to model hourly electricity demand for a large cross-section of $N$ providers or regions. In this scenario, each series $y_{i\tau}$ has two integer-valued seasonalities: \emph{Intra-week} (7 days), corresponding to $s_1 \in \{1,\dots,7\}$;  \emph{Intra-day} (24 hours), corresponding to $s_2 \in \{1,\dots,24\}$.
Hence, we set $M=2$. With $\mathcal{T}$ total hourly observations, each complete week corresponds to $7 \times 24 = 168$ hours, so $T=\frac{\mathcal{T}}{168}$. In this setting, we re-index each observation $\tau$ by $(s_1,s_2,t)$, so that
\[
y_{i,s_1,s_2,t}, 
\quad i=1,\dots,N,\;\;
s_1=1,\dots,7,\;\; s_2=1,\dots,24,\;\; t=1,\dots,T.
\]
A two-layer hierarchical factor model can then be written as
\begin{equation}
	\begin{array}{rll}
		y_{i,s_1,s_2,t} &= \mu_{i,s_1,s_2} + \sigma_{i,s_1,s_2} \epsilon_{i,s_1,s_2,t} \\[0.1in]
		 \epsilon_{i,s_1,s_2,t} &= \sum_{r=1}^{R} {\lambda}_{i,r} {f}^{(2)}_{r,s_1,s_2,t} +\; \nu_{i,s_1,s_2,t} \\[0.1in]
		{f}^{(2)}_{r,s_1,s_2,t} &= \sum_{k_2=1}^{K_2}{\beta}^{(2)}_{s_2,k_2} f^{(1)}_{r,s_1,k_2,t} + \eta^{(2)}_{r,s_1,s_2,t} \\[0.1in]
		{f}^{(1)}_{r,s_1,k_2,t} &= \sum_{k_1=1}^{K_1}\beta^{(1)}_{s_1,k_1}\,f_{r,k_1,k_2,t} + \eta^{(1)}_{r,s_1,k_2,t} 
	\end{array}
	\label{eq:electricity_model}
\end{equation}
where $\mu_{i,s_1,s_2,t}$ and $\sigma_{i,s_1,s_2,t}$ varies by provider $i$, day of the week $s_1$ and hour of the day $s_2$, capturing differences in location and scale, respectively,  $\mathbf{f}^{(2)}_{s_1,s_2,t}$ is a $R$-dimensional vector capturing the latent factor structure across individuals in week $t$. For each $r$-th layer, $\mathbf{f}^{(1)}_{r,s_1,\cdot,t}$ is a $K_2$-dimensional vector capturing the \emph{hourly} (intra-day) latent factor structure within day $s_1$ of week $t$, and for each $k_2$-th intra-day layer, $\mathbf{f}_{r,\cdot,k_2,t}$ is a $K_1$-dimensional vector capturing the \emph{daily} (intra-week) factor structure within week $t$. Each element $f_{r,k_1,k_2,t}$ may be interpreted as an overarching time-dependent factor, e.g., a baseline weekly trend. The loading matrices $\boldsymbol{\Lambda}$, $\boldsymbol{B}^{(2)}$, and $\boldsymbol{B}^{(1)}$ specify how each provider $i$ responds to the lower-level factors $\mathbf{f}^{(2)}_{s_1,s_2,t}$ , and how different hours and days modify higher-level factors, $f_{r,k_1,k_2,t} $. The random shocks $\epsilon_{i,s_1,s_2,t}$, $\eta^{(2)}_{r,s_1,s_2,t}$, and $\eta^{(1)}_{r,s_1,k_2,t} $ account for unexplained variation or measurement noise. By disentangling \emph{intra-day} and \emph{intra-week} seasonalities in a multi-level fashion, this specification accommodates the common cycles observed in electricity usage (e.g., higher load during daytime hours, differences between weekdays and weekends), while also allowing for cross-sectional heterogeneity across providers $i$. 

\subsection{Estimation}
Section~\ref{sec:general_framework} proposes rearranging the original panel time series dataset of dimension \(N \times \mathcal{T}\) into a sequence of \(T\) multidimensional arrays (tensors) of dimension \( N \times S_1 \times \cdots \times S_M. \) Once expressed in its compact form~\eqref{eq:gen_frame_comp}, the multi-level factor model in~\eqref{eq:gen_frame} naturally fits naturally fits into the broader class of tensor factor models~\citep{Chen2022, Barigozzi2023b}, allowing one to leverage existing estimation methods for the loading matrices \(\boldsymbol{\Lambda}, \mathbf{B}^{(1)}, \dots, \mathbf{B}^{(M)}\) and the latent factor tensors \(\mathcal{F}_t = [f_{k,k_1,\dots,k_M,t}]\).

We briefly overview some of the elements of tensor notation considered. A \(K\)-dimensional tensor is an array \(\mathcal{X} = [x_{i_1,\dots,i_K}] \in \mathbb{R}^{p_1 \times \cdots \times p_K}\) whose entries \(x_{i_1,\dots,i_K}\) are indexed by \(K\) coordinates.  Each dimension of \(\mathcal{X}\) is referred to as a \emph{mode}, so the \(k\)-th dimension is called \emph{mode-\(k\)}. A \emph{mode-\(k\) fiber} is obtained by fixing all indices except the one in mode \(k\), resulting in a 1-dimensional slice, that is, the $p_k$-dimensional vector $\bigl(\,\mathcal{X}_{i_1,\dots, i_{k-1}, \,\cdot,\,i_{k+1},\dots, i_K}\bigr)$.  
\emph{Mode-\(k\) matricization} (also called unfolding)), denoted $\text{mat}_{(k)} (\mathcal{X}_t)$, maps \(\mathcal{X}\) to a matrix \(\mathbf{X}_{(k)} \in \mathbb{R}^{p_k \times (p_1 \cdots p_{k-1} \, p_{k+1} \cdots p_K)}\) by setting mode~\(k\) as the row dimension and merging all other modes as columns. Formally, each entry \(\mathcal{X}_{i_1,\dots,i_K}\) becomes \(\mathbf{X}_{(k)}(i_k, j)\), where \(j\) encodes \((i_1,\dots,i_{k-1}, i_{k+1},\dots,i_K)\). For a matrix \(\mathbf{A} \in \mathbb{R}^{d \times p_k}\), the \emph{mode-\(k\) product} \(\mathcal{Y} = \mathcal{X} \times_k \mathbf{A}\) is defined by multiplying each mode-\(k\) fiber of \(\mathcal{X}\) by \(\mathbf{A}\). In the resulting tensor \(\mathcal{Y}\), the size of mode~\(k\) changes from \(p_k\) to \(d\), while all other modes remain unchanged.

Let $\mathcal{X}_t = [x_{i,\mathbf{s},t}]$ be a three-dimensional tensor in $\mathbb{R}^{N\times S_1 \times \cdots \times S_M}$ where
\[
x_{i,\mathbf{s},t} = \frac{y_{i,\mathbf{s},t}-\mu_{i,\mathbf{s}}}{{\sigma}_{i,\mathbf{s}}}.
\] 
The multi-level factor model in \eqref{eq:gen_frame}, written in compact form in \eqref{eq:gen_frame_comp}, can be expressed as
\begin{equation}
	\mathcal{X}_t = \mathcal{F}_t \times_{1} \boldsymbol{\Lambda} \times_2 \textbf{B}^{(1)} \cdots \times_{M+1} \textbf{B}^{(M)} + \mathcal{E}_t
	\label{eq:tensor_factor_model}
\end{equation}
where $\mathcal{E}_t=[\varepsilon_{i,\mathbf{s},t}] \in \mathbb{R}^{N\times S_1 \times \cdots \times S_M}$ and $\mathcal{F}_t=[f_{r,\dots, k_J, \dots, t}]  \subset \mathbb{R}^{R \times K_1 \times \cdots \times K_M}$.  Using the unfolding operation, one obtains the following $M+1$ equations,
\begin{equation}
	\begin{split}
		\text{mat}_{(1)}{\mathcal{X}}_t & = \text{mat}_{(1)} \Big(\mathcal{F}_t \times_{1} \boldsymbol{\Lambda} \times_2 \textbf{B}^{(1)} \cdots \times_{M+1} \textbf{B}^{(M)} + \mathcal{E}_t\Big) \\
		& = {\boldsymbol{\Lambda}} \text{mat}_{(1)} \Big(\mathcal{F}_t \Big) \mathbf{B}' + \text{mat}_{(1)} \Big(\mathcal{E}_t\Big),
		\label{eq:tensor_scalar_lambda}
	\end{split}
\end{equation}
with $\mathbf{B}=\mathbf{B}^{(M)} \otimes \cdots \otimes \mathbf{B}^{(1)}$, and, for each $j=1,\dots,M$,
\begin{equation}
	\begin{split}
		\text{mat}_{(j+1)}{\mathcal{X}}_t & = \text{mat}_{(j+1)} \Big(\mathcal{F}_t \times_{1} \boldsymbol{\Lambda} \times_2 \textbf{B}^{(1)} \cdots \times_{M+1} \textbf{B}^{(M)} + \mathcal{E}_t\Big) \\
		& = \textbf{B}^{(j)}  \text{mat}_{(j+1)} \Big(\mathcal{F}_t \Big) \boldsymbol{\Gamma}^{(j)\prime} + \text{mat}_{(j+1)} \Big(\mathcal{E}_t\Big),
		\label{eq:tensor_scalar_b}
	\end{split}
\end{equation}
with $\boldsymbol{\Gamma}^{(j)}= \mathbf{B}^{(M)} \otimes \cdots \otimes \mathbf{B}^{(j+1)} \otimes \mathbf{B}^{(j-1)} \otimes \cdots \otimes \mathbf{B}^{(1)} \otimes \boldsymbol{\Lambda}$. We define the following estimators for the location and scale parameters, $\widetilde{\mu}_{i,\mathbf{s}}=\frac{1}{T}\sum_{t=1}^{T} y_{i,\mathbf{s},t}$ and $\widetilde{\sigma}_{i,\mathbf{s}}=\frac{1}{T}\sum_{t=1}^{T} (y_{i,\mathbf{s},t}-\widehat{\mu}_{i,\mathbf{s}})^2$.  Estimators $\widetilde{\boldsymbol{\Lambda}}, \widetilde{\mathbf{B}}^{(1)}, \cdots, \widetilde{\mathbf{B}}^{(M)}$ can then be obtained with the projection algorithm of \cite{Yu2022} and detailed in Algorithm \ref{algo:estimators}.  Finally, an estimator of the common factor tensor is obtained by linear projection as
\[
\widetilde{\mathcal{F}}_t = \frac{1}{NS} \, \mathcal{X}_t \times^M_{j=1} \widetilde{\mathbf{B}}^{(j)}\times_{M+1} \widetilde{\boldsymbol{\Lambda}}.
\]
and the fitted values as
\begin{equation}
\widetilde{\mathcal{Y}}_t = \widetilde{\mathcal{M}} + \widetilde{\mathcal{S}} \odot \left(\widetilde{\mathcal{F}}_t \times_{1} \widetilde{\boldsymbol{\Lambda}} \times_2 \widetilde{\textbf{B}}^{(1)} \cdots \times_{M+1} \widetilde{\textbf{B}}^{(M)} \right)
\label{eq:fitted}
\end{equation}
where $\widetilde{\mathcal{M}}=[\widehat{\mu}_{i,\mathbf{s}}]$, $\widetilde{\mathcal{S}}=[\widehat{\sigma}_{i,\mathbf{s}}]$ and $\odot$ denotes the element-wise (Hadamard) product.

\begin{algorithm}
	\caption{Projected Method for Estimating the Loading Spaces}
	\KwIn{Tensor data $\{\mathcal{X}_t, 1 \leq t \leq T\}$, factor numbers $R,K_1, \dots, K_M$}
	\vspace{0.5em} 
	\KwOut{Factor loading matrices $\widetilde{\boldsymbol{\Lambda}}$ and $\{\widetilde{\mathbf{B}}^{(j)}, 1 \leq j \leq M\}$}
	Compute initial estimators $\widehat{\mathbf{B}}$ and $\{\widehat{\boldsymbol{\Gamma}}^{(j)}, 1 \leq j \leq M\}$ as 
	\[
	\begin{array}{rll}
		\widehat{\mathbf{B}} &=& \sqrt{S} \times \text{the $R$ eigenvectors of } \frac{1}{TNS}\sum\limits_{t=1}^{T} \mathbf{X}'_{1,t} \mathbf{X}_{1,t}, \\[0.1in]
		\widehat{\boldsymbol{\Gamma}}^{(j)} &=& \sqrt{NS_{-j}} \times \text{the $(RK_{-j})$ eigenvectors of } \frac{1}{TNS}\sum\limits_{t=1}^{T} \mathbf{X}'_{j+1,t} \mathbf{X}_{j+1,t}, \quad 1 \leq j \leq M. \\[0.1in]
	\end{array}
	\]
	where $\mathbf{X}_{\bullet,t} = \text{mat}_{(\bullet)}{\mathcal{X}}_t$, $S_{-j}=\prod_{i\not= j}^{M}S_i$ and $K_{-j}=\prod_{i\not= j}^{M}K_i$.
	
	Compute projected estimators $\widetilde{\boldsymbol{\Lambda}}$ and $\{\widetilde{\mathbf{B}}^{(j)}, 1 \leq j \leq M\}$ as
	\[
	\begin{array}{rll}
		\widetilde{\boldsymbol{\Lambda}} &=& \sqrt{N} \times \text{the $R$ eigenvectors of } \frac{1}{TNS}\sum\limits_{t=1}^{T} \mathbf{X}_{1,t} \frac{\widehat{\mathbf{B}}'\widehat{\mathbf{B}}}{S}  \mathbf{X}'_{1,t}, \\[0.1in]
		\widetilde{\mathbf{B}}^{(j)} &=& \sqrt{S_{j}} \times \text{the $(K_{j})$ eigenvectors of } \frac{1}{TNS}\sum\limits_{t=1}^{T} \mathbf{X}_{j+1,t} \frac{\widehat{\boldsymbol{\Gamma}}^{(j)\prime}\widehat{\boldsymbol{\Gamma}}^{(j)}}{S_{-j}}  \mathbf{X}'_{j+1,t}, \quad 1 \leq j \leq M. \\[0.1in]
	\end{array}
	\]
	\label{algo:estimators}
\end{algorithm}

\paragraph{Remark.} Note that if the process in \eqref{eq:tensor_factor_model} satisfies Assumptions 1-3 in \cite{Yu2022}, there exists matrices $\widehat{\mathbf{H}}$ and $\widehat{\mathbf{J}}_j$ satisfying $\widehat{\mathbf{H}}'\widehat{\mathbf{H}}\xrightarrow{p}\mathbb{I}_{R}$ and $\widehat{\mathbf{J}}'_j\widehat{\mathbf{J}}_j\xrightarrow{p}\mathbb{I}_{K_j}$, with $j=1,\dots,M$, such that,
\[
\begin{array}{rll}
\frac{1}{\sqrt{N}} \left\lVert \widetilde{\boldsymbol{\Lambda}} - \boldsymbol{\Lambda}\widehat{\mathbf{H}}  \right\rVert &=& O_p\left(\min\left\{\frac{1}{\sqrt{T}}, \frac{1}{N}\right\}\right) \\[0.1in]
\frac{1}{\sqrt{S_j}} \left\lVert \widetilde{\mathbf{B}}^{(j)} - \mathbf{B}\widehat{\mathbf{J}}_j  \right\rVert &=& O_p\left(\min\left\{\frac{1}{\sqrt{TN}}, \frac{1}{T}, \frac{1}{N}\right\}\right)
\end{array}
\]
as $\min\left\{N,T\right\} \rightarrow\infty$.

\subsection{Forecasting}

In line with classical approaches to time series analysis and forecasting, the estimated common factors are modeled within a standard time-series framework. Once the multi-level factor model has been estimated, the temporal dynamics of the factors $\mathcal{F}_t$ can be separately analyzed using well-established univariate or multivariate time series models, such as autoregressive integrated moving average (ARIMA) models. Given their central role in capturing dependencies over time, such models allow for extrapolating future values of the factor tensor $\mathcal{F}_t$ and, subsequently, the observed data. Notice that the factors might display (\emph{non-integer}) seasonalities therefore, one might consider some deseasonalization technique before using ARIMA and then add the estimated seasonal component back to the forecast obtained from ARIMA. Alternatively, one can directly consider an ARIMA model with deterministic seasonality or a seasonal ARIMA (SARIMA) approach, which explicitly incorporates seasonal components in the forecasting process.

Once the factor dynamics have been estimated, future values of the factor tensor, denoted as $\widetilde{\mathcal{F}}_{T+n|T}$, are obtained through recursive forecasting. Given the predicted $n$-step-ahead factor tensor, forecasts of the original observed data $\mathcal{Y}_t$ are obtained by plugging $\widetilde{\mathcal{F}}_{T+n|T}$ into the estimated factor model structure. Specifically, the $n$-step-ahead forecast is computed as
\begin{equation}
	\widetilde{\mathcal{Y}}_{T+n|T} = \widetilde{\mathcal{M}} + \widetilde{\mathcal{S}} \odot \left(\widetilde{\mathcal{F}}_{T+n|T} \times_{1} \widetilde{\boldsymbol{\Lambda}} \times_2 \widetilde{\mathbf{B}}^{(1)} \cdots \times_{M+1} \widetilde{\mathbf{B}}^{(M)} \right).
	\label{eq:forecast}
\end{equation}
Here, $\widetilde{\mathcal{M}}$ and $\widetilde{\mathcal{S}}$ ensure that forecasts are properly rescaled to match the original data scale, and the factor-loading structure is used to reconstruct the forecasts in the multidimensional tensor space.

\section{High-dimensional modeling and forecasting of electricity demand}
\label{sec:empirical}
This section applies the model presented in Section \ref{sec:hourly_model} to the PJM dataset discussed in Section \ref{sec:data}. First, the original panel time series dataset of hourly observations with dimension $N=9$ and $\mathcal{T}=57,456$, is folded into a weakly time series ($T=342$) of tensor with dimension $9\times 24 \times 7$. Section \ref{sec:modelfit} provides an assessment of the model fit evaluating its ability to replicate the main empirical aspects discussed in Section \ref{sec:data}. Then, we split the tensor time series into two halves to obtain a training and the test sample of equal length ($T_{\text{train}}=T_{\text{test}}=171$), and evaluate the predictive ability of the model in Section \ref{sec:forecast}.

\subsection{Model fit assessment}
\label{sec:modelfit}

The exploratory data analysis presented in Section~\ref{sec:data} highlights several key aspects to address when modeling hourly electricity demand across different companies. First, the empirical mean and standard deviation vary by company, day of the week, and hour of the day; hence, these differences are captured by the location and scale parameters $\mu_{i,\mathbf{s}}$ and $\sigma_{i,\mathbf{s}}$. Second, the companies' dynamics are highly similar, suggesting that a single factor may suffice to capture cross-sectional dependencies in the company dimension. Third, the weekly seasonal pattern is also similar across providers and remains stable throughout the year, indicating that only one factor may be needed for the weekly dimension. Fourth, the daily (intra-day) seasonal pattern changes markedly with the season, implying that two factors may be necessary to capture its evolution over the year.

\begin{figure}[htpb]
	\centering  
	\includegraphics[scale=0.90]{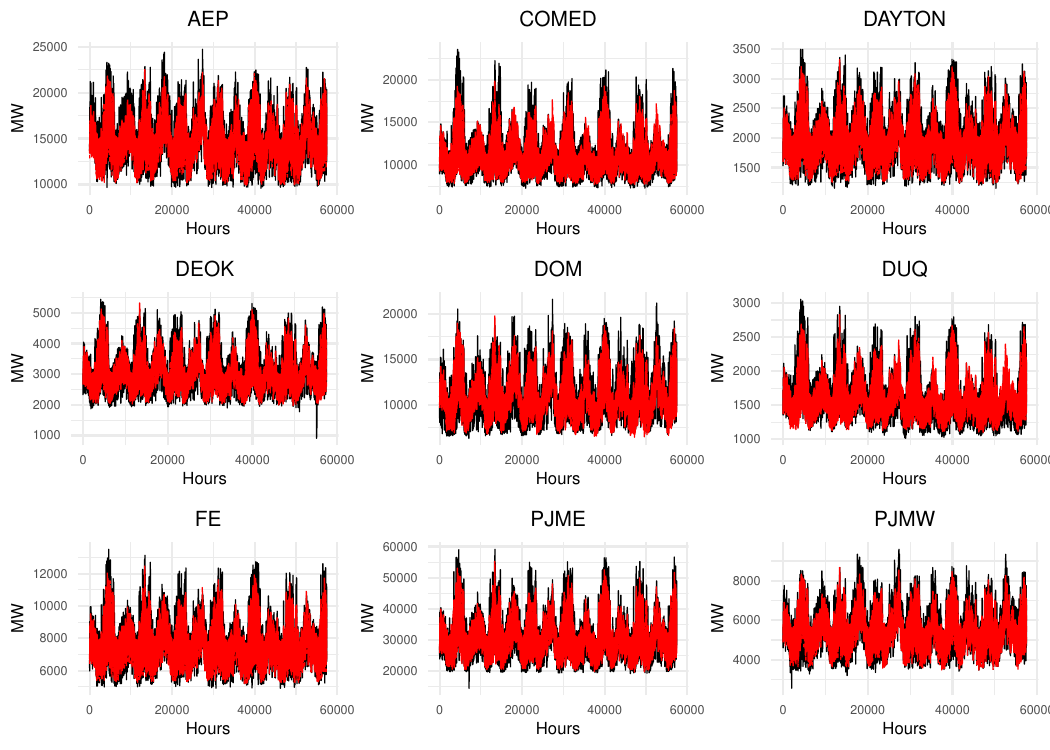} 
	\caption{Time series of hourly energy demand in Mega Watts (black) and fitted values from the two-factor model ($k=1, k_1=1, k_2=2$).} 
	\label{fig:TimeSeriesPlot_fitted_two_factor} 
\end{figure}

\begin{figure}[htpb]
	\centering
	\includegraphics[width=0.8\textwidth]{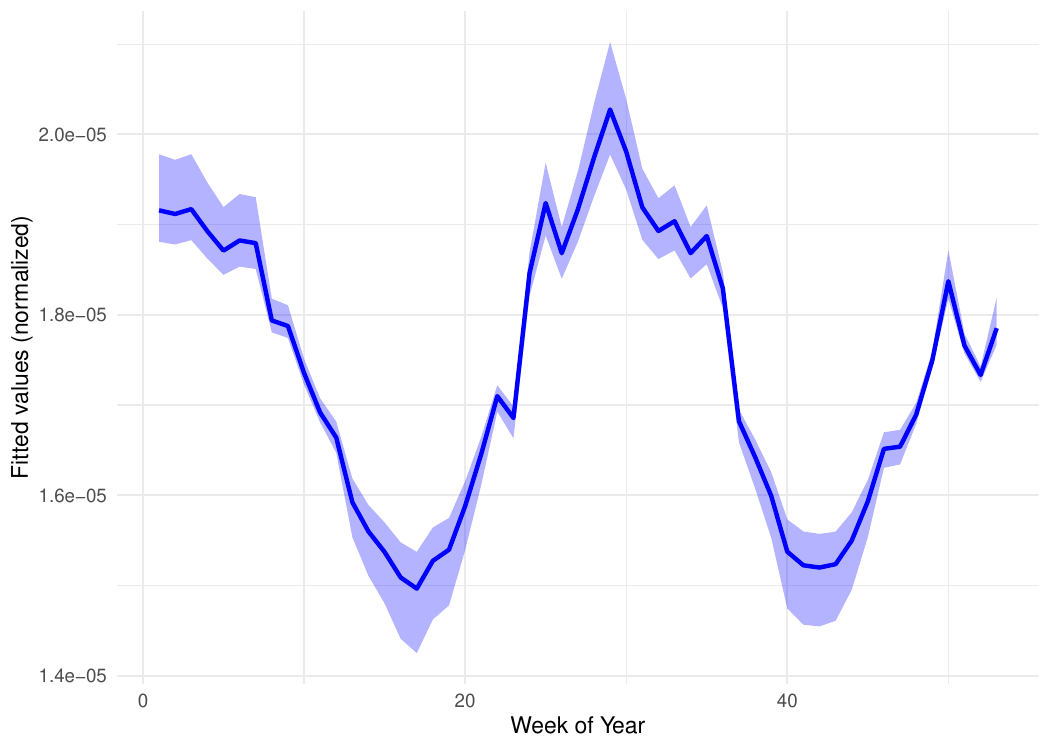}
	\caption{Weakly average (normalized) fitted values from the two-factor model ($k=1, k_1=1, k_2=2$) over the year. Shaded areas and solid lines display the range and the average across companies, respectively.}
	\label{fig:IntraYearSeasonality_fitted_two_factor}
\end{figure}

\begin{figure}[htpb]
	\centering
	\includegraphics[width=0.8\textwidth]{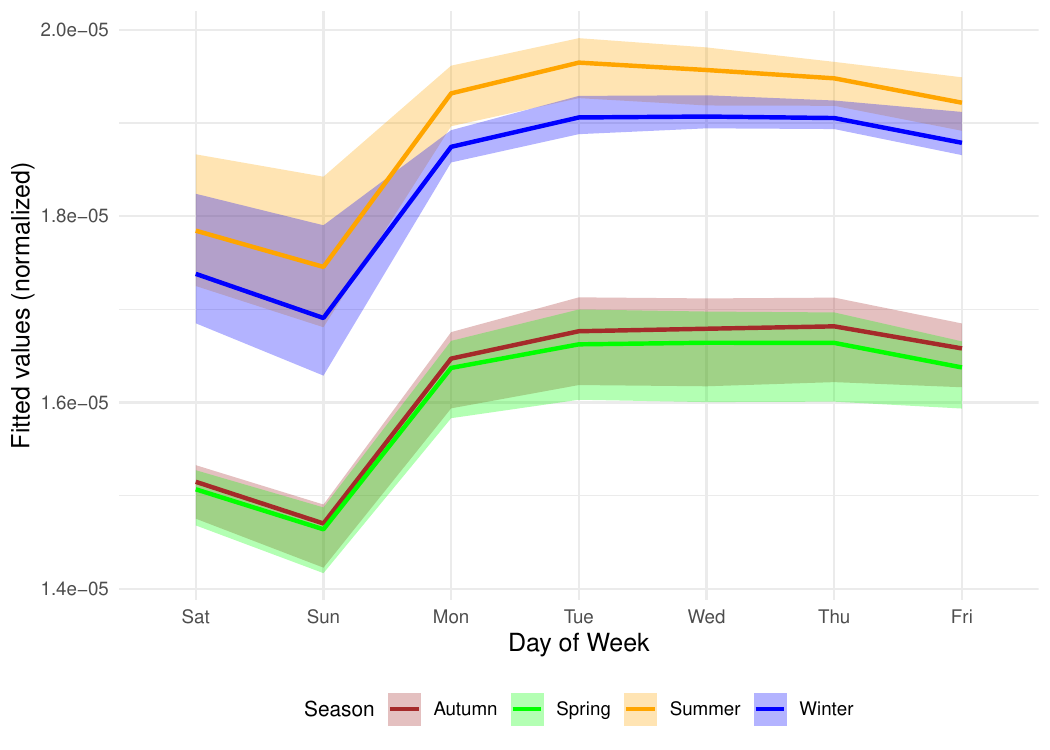}
	\caption{Average (normalized) fitted values from the two-factor model ($k=1, k_1=1, k_2=2$) for every day of the week in each season. Shaded areas and solid lines display the range and the average across companies, respectively.}
	\label{fig:IntraWeekSeasonality_fitted_two_factor}
\end{figure}

\begin{figure}[ht]
	\centering
	\begin{subfigure}[b]{0.45\textwidth}
		\centering
		\includegraphics[width=\textwidth]{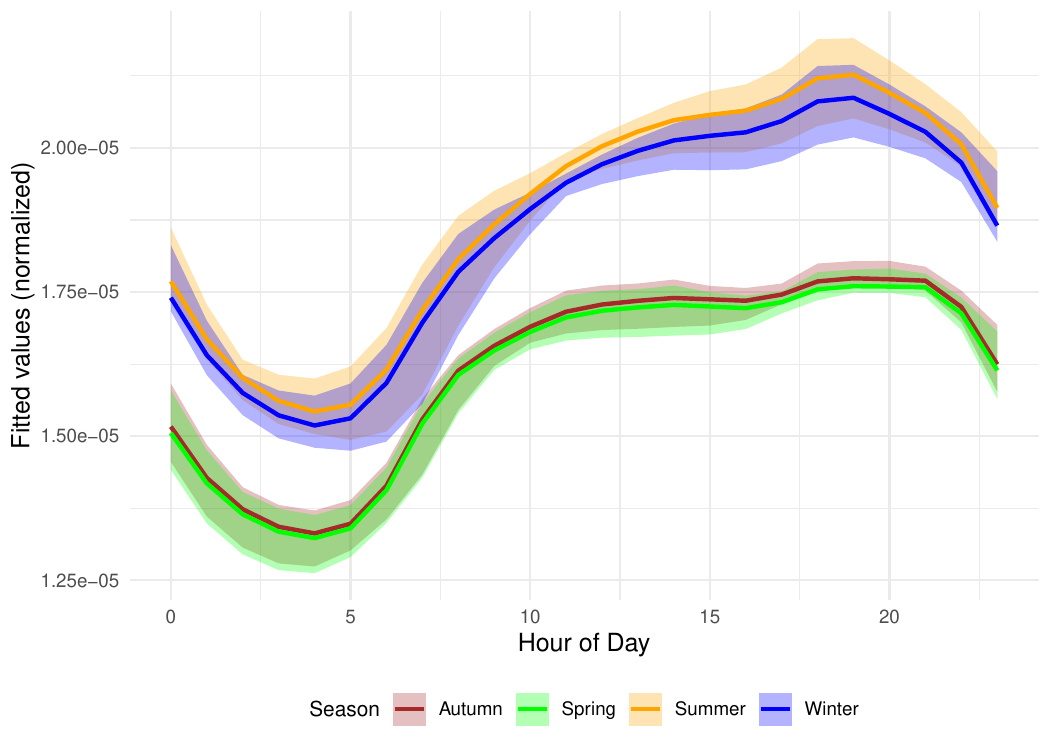}
		\caption{One-factor}
		\label{fig:plot1}
	\end{subfigure}
	\begin{subfigure}[b]{0.45\textwidth}
		\centering
		\includegraphics[width=\textwidth]{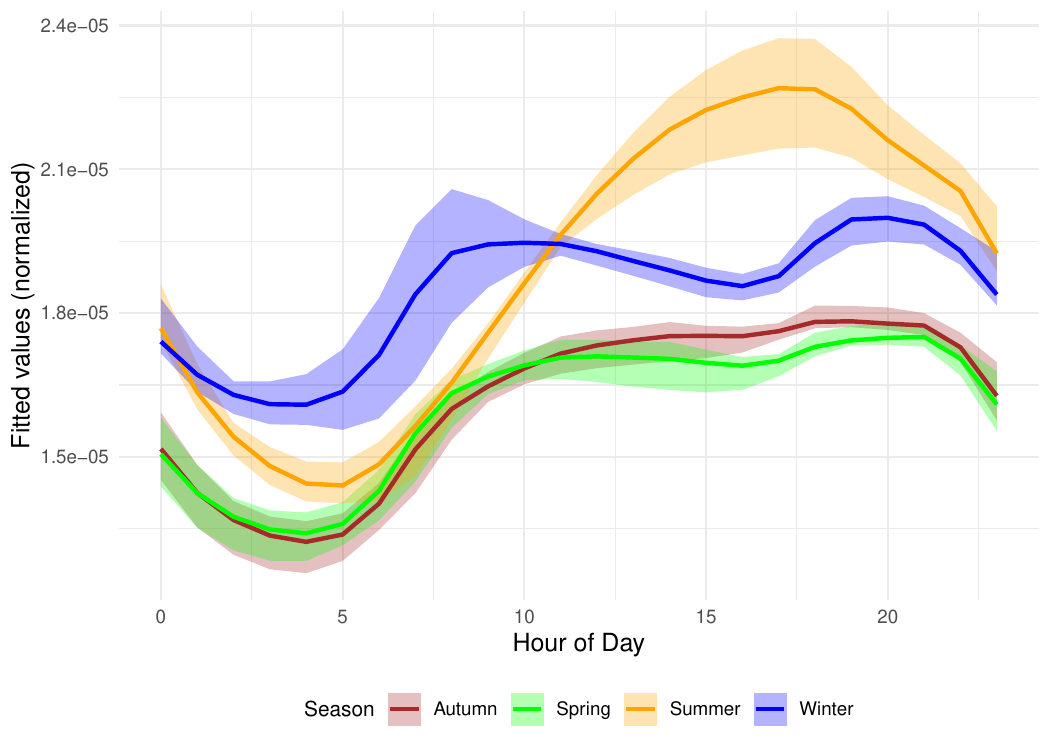}
		\caption{Two-factor}
		\label{fig:plot2}
	\end{subfigure}
	\caption{Average (normalized) fitted values from the two-factor model ($k=1, k_1=1, k_2=2$) for every hour of the day in each season. Shaded areas and solid lines display the range and the average across companies, respectively.}
	\label{fig:IntraDaySeasonality_fitted}
\end{figure}

Guided by these considerations, we explore two specifications of the model in \eqref{eq:electricity_model}. The first is a parsimonious version with one global factor driving the entire system ($k=1, k_1=1, k_2=1$). The second is a richer specification that accommodates possible variations in the intra-day seasonal pattern ($k=1, k_1=1, k_2=2$). To inform our choice, we apply the eigenvalue-ratio criterion described in Section~2.3 of \cite{Barigozzi2024}, which favors the more parsimonious model. Nonetheless, empirical results show that the richer specification consistently achieves a better fit in terms of mean squared error (MSE). Figure~\ref{fig:TimeSeriesPlot_fitted_two_factor} displays the fitted values from the two-factor model, obtained via \eqref{eq:fitted}. To highlight how well this model reproduces the salient patterns in the data, Figure~\ref{fig:IntraYearSeasonality_fitted_two_factor} shows the weekly average (normalized) fitted values over the year, while Figure~\ref{fig:IntraWeekSeasonality_fitted_two_factor} plots the average (normalized) fitted values by day of the week, broken down by season. These patterns align closely with those observed in Figures~\ref{fig:IntraYearSeasonality} and~\ref{fig:IntraWeekSeasonality}, underscoring the good fit of our model. Although analogous figures for the one-factor model appear similarly robust, we omit them for brevity. Finally, Figure~\ref{fig:IntraDaySeasonality_fitted} compares the intra-day seasonal patterns from the one-factor and two-factor models. In relation to the observed data, the two-factor specification (Panel~(b)) more effectively captures the complexity of hourly electricity usage across different seasons, underscoring the benefit of allowing for additional factors at the daily level.

\begin{figure}[ht]
	\centering
	\includegraphics[width=0.8\textwidth]{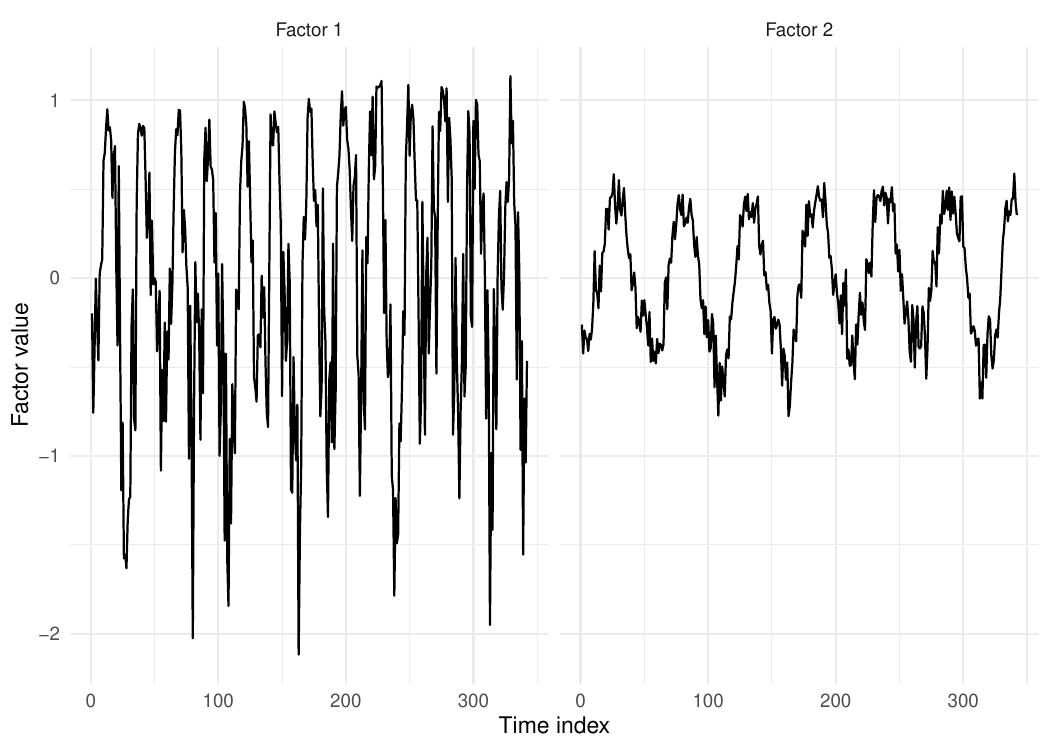}
	\caption{Time series of estimated factors from the two-factor model ($k=1, k_1=1, k_2=2$).} 
	\label{fig:TimeSeriesPlot_estimated_two_factor} 
\end{figure}

\begin{figure}[ht]
	\centering
	\includegraphics[width=0.8\textwidth]{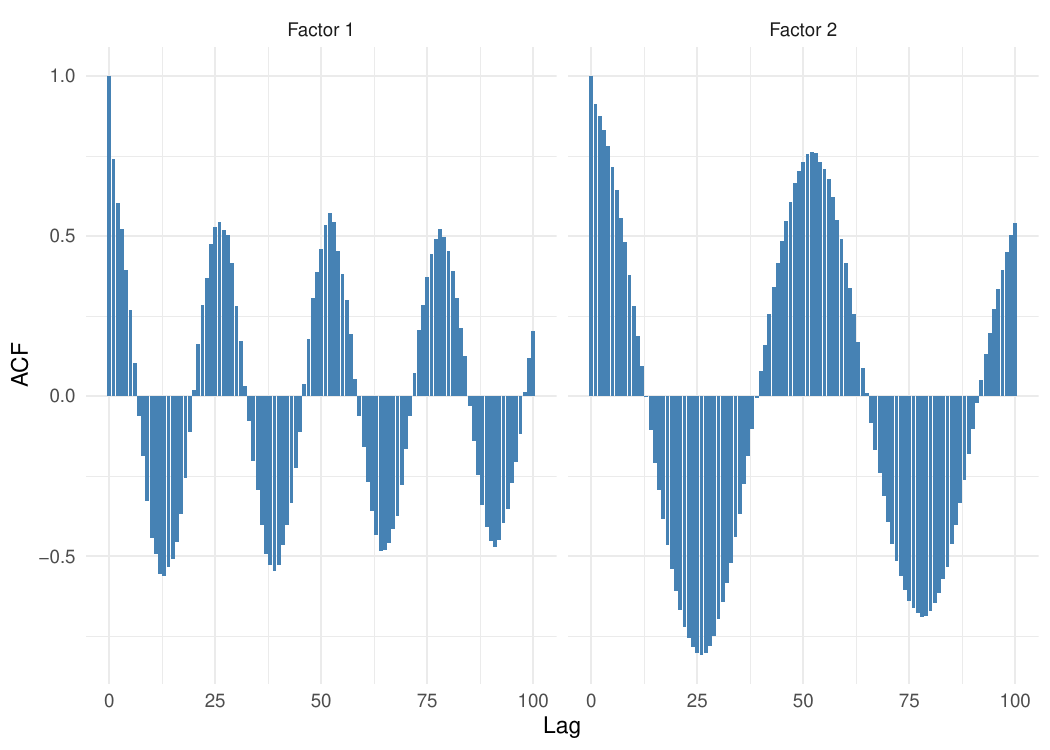}
	\caption{Sample autocorrelation of the estimated factors from the two-factor model ($k=1, k_1=1, k_2=2$).} 
	\label{fig:ACF_estimated_two_factor} 
\end{figure}

Based on these findings, we proceed with the forecast evaluation using the two-factor specification as our benchmark model. Forecasting $\mathcal{Y}_t$ requires predicting the future values of the factor tensor $\mathcal{F}_t$. While factor loadings remain constant over time, the factors themselves capture the underlying dynamics that drive the observed variables. Therefore, before implementing the forecasting framework, we first examine the time series properties of the estimated factors. Figures \autoref{fig:TimeSeriesPlot_estimated_two_factor} and \autoref{fig:ACF_estimated_two_factor} illustrate the time series evolution and sample autocorrelation functions of the extracted factors. Both series exhibit pronounced seasonal patterns. The first factor displays an average cycle of approximately 26 weeks, reflecting peak demand periods in summer and winter, as highlighted in \autoref{fig:IntraYearSeasonality}. The second factor follows a strong annual cycle, peaking in summer, which allows for greater flexibility in capturing variations in diurnal patterns, as shown in \autoref{fig:IntraDaySeasonality}. These seasonal characteristics are crucial for accurately modeling and forecasting the factor time series. Properly accounting for them enhances forecast precision and ensures a more reliable representation of the underlying structure of the data.

\subsection{Forecast evaluation}
\label{sec:forecast}
We employ a rolling window approach to generate pseudo-out-of-sample forecasts. The window begins at $t_0 = 1$, extends to $t = T_{\text{test}}$, and advances by one period at each iteration. Models are re-estimated in each window and produce $n$-step-ahead forecasts. We consider forecast horizons of $n = \{1, 4, 13, 26\}$, corresponding to one week, one month, one quarter, and one semester ahead. The original sample is split into two halves, yielding $T_{\text{test}} = 171$.

We perform a separate evaluation of the forecasts on each electricity provider. Let $\mathbf{Y}_{i,t}$ denote the $7\times 24$ matrix of observations corresponding to the $i$th provider at time $t$ and , to evaluate the $n$-step ahead forecasts we consider the following MSE measure:
\begin{equation}
	MSE_{n} = \frac{1}{W \times 7 \times 24} \mathlarger{\sum^{W}_{w=1}} \norm{\mathbf{Y}_{i,T_{\text{test}} +w+n} - \widetilde{\mathbf{Y}}_{i,T_{\text{test}}+w+n|T_{\text{test}}+w}}^{2}_F
	\label{eq:MSE}
\end{equation}
where $w$ is the window of reference, $n$ is the horizon of the forecast, $W=T-n$ is the total number of rolling windows and $\widetilde{\mathbf{Y}}_{i,t+n|t}$ is the $n$-step ahead forecast.

Using our tensor factor model, we obtain forecasts $\widetilde{\mathbf{Y}}_{i,T_{\text{test}}+w+n|T_{\text{test}}+w}$ for each window and provider by extracting them from the forecasted tensor computed according to Equation \eqref{eq:forecast}. The core of our forecasting procedure involves fitting a time series model to each factor score series within every rolling window. Figures \ref{fig:TimeSeriesPlot_estimated_two_factor} and \ref{fig:ACF_estimated_two_factor} indicate strong seasonality in the factor series, which must be accounted for in the forecasting process. To address this, we apply seasonal adjustment by removing a deterministic seasonal component from each factor series using the classical additive decomposition \citep{Hyndman2018}. Specifically, we subtract this estimated seasonal component before fitting a time series model. To capture the remaining dynamics, we fit a simple autoregressive model of order 1 (AR(1)) to each seasonally adjusted factor series. Once forecasts are generated, we restore seasonality by adding back the previously removed seasonal component to the predicted scores. Since the seasonal pattern is assumed to be deterministic and constant, incorporating the future seasonal component is straightforward. 

To evaluate the performance of our model, we compare it against the following alternative approaches, computed for each rolling window:
\begin{itemize}
	\item \textbf{Matrix factor model}.  Instead of modeling the entire tensor time series $\{\mathcal{Y}_t\}$, we model the time series of $7 \times 24$ matrices, $\{\mathbf{Y}_{i,t}\}$, for each provider. Specifically, we adopt a matrix factor model with: one factor for the day-of-the-week dimension; two factors for the hour-of-the-day dimension. Factor matrices and loadings are estimated using the procedure of \cite{Yu2022}. Forecasts of each factor are obtained using an AR(1) model fitted to the seasonally adjusted factor series. The final forecasts for $\{\mathbf{Y}_{i,t}\}$ are reconstructed by combining the predicted factors with the estimated loadings.
	\item \textbf{Vector factor model}. Matrix and tensor time series factor models can be interpreted as constrained versions of high-dimensional vector factor models (VFMs), as discussed in \cite{Yu2022}. To benchmark our approach, we apply a VFM to the vectorized provider time series $\{\mathbf{Y}_{i,t}\}$. The estimation follows a similar strategy to our tensor approach: factors and factor loadings are extracted using PCA; an AR(1) model is fitted to the seasonally adjusted factor series to forecast the factors.
	\item \textbf{Functional time series model}. We implement the functional time series (FTS) approach of \citet{Shang_2012}, applied separately to each provider’s vectorized time series $\{\mathbf{Y}_{i,t}\}$. The procedure follows these steps:
	\begin{enumerate}
		\item The original vector is divided into seven daily vectors, each containing hourly electricity demand observations for a specific day of the week across all weeks in the dataset.
		\item Each daily vector is reshaped into a matrix, where rows correspond to the 24 hours of the day and columns represent weeks in the rolling window.
		\item Using the R package \texttt{FTS}, we apply smoothing splines to convert columns into hourly curves. FPCA is then used to decompose the curves into functional principal components and their associated time series of component scores.
		\item An ARIMA model is fitted to the component scores to generate $n$-step-ahead forecasts. The forecasts for each day matrix are reconstructed by combining the predicted scores with the functional principal components.
		\item The forecasted daily matrices are merged to form the final provider matrix forecast $\{\mathbf{Y}_{i,t}\}$.
	\end{enumerate}
\end{itemize}
These alternative models serve as benchmarks, allowing us to assess the relative performance of our proposed approach.

Table \ref{tab:mse_oos_results} presents the out-of-sample relative mean squared error (MSE) for each forecasting model across different providers and forecasting horizons. The relative MSE is computed as the MSE in \eqref{eq:MSE} divided by the average standard deviation of the out-of-sample observations within each rolling window. This normalization accounts for variations in the scale of electricity demand across different time periods, allowing for a more meaningful comparison of forecasting accuracy.

The results clearly indicate that our tensor factor model achieves the lowest relative MSE across most providers and forecasting horizons, demonstrating its superior forecasting performance. By explicitly modeling the multidimensional structure of the data, the tensor factor approach effectively captures complex seasonal patterns and long-term dependencies, leading to more accurate forecasts. 

For short-term forecasting (week-ahead), the tensor factor model performs best for three providers (DAYTON, DEOK, and DOM), while the matrix factor model achieves the lowest MSE for the remaining providers. The performance gap between the two models is relatively small, suggesting that both approaches are effective for short-term forecasting. However, as the forecasting horizon extends, the advantage of the tensor factor model becomes more pronounced. In the month-ahead and quarter-ahead forecasts, our tensor approach consistently outperforms other approaches, achieving the lowest relative MSE for a majority of providers, including COMED, DEOK, DUQ, PJME, and PJMW. For long-term forecasting (semester-ahead), the tensor factor model continues to exhibit superior performance, maintaining the lowest relative MSE for five providers (AEP, COMED, DEOK, DUQ, and PJMW). In contrast, the matrix factor model performs best only for DOM, while the vector factor model achieves the lowest error for DAYTON. The fact that the tensor factor model maintains its advantage even in longer forecasting horizons suggests that it effectively mitigates the accumulation of forecasting errors over time, a common challenge in electricity demand prediction. Comparatively, the matrix factor model provides competitive performance in short-term forecasting but is generally outperformed by the tensor factor model in longer horizons. The vector factor model struggles to match the accuracy of tensor-based approaches, as it does not fully exploit the multidimensional dependencies in the data. The functional time series model consistently exhibits the highest relative MSE across all horizons, indicating that it is less effective at capturing the intricate temporal structures required for accurate forecasting.

Overall, these results highlight the effectiveness of our tensor factor model as the most reliable forecasting approach. By leveraging the inherent structure of the data, this model consistently outperforms alternative methods across different forecasting horizons and providers. The findings strongly support the adoption of tensor-based methodologies in electricity demand forecasting, particularly when long-term accuracy and robustness are critical.

	\begin{table}[ht]
		\centering
			\begin{tabular}{lccccccccc}
				\toprule
				\textbf{Week Ahead} & AEP & COMED & DAYTON & DEOK & DOM & DUQ & FE & PJME & PJMW \\
				\midrule
				\multicolumn{10}{c}{\textbf{\textbf{Tensor factor model}}} \\
				\midrule				
				Week  & 0.5803 & 0.5929 & \bf{0.5668} & \bf{0.5971} & \bf{0.6173} & 0.6152 & \bf{0.5658} & 0.5576 & 0.6009 \\
				Month & 0.6148 & \bf{0.6191} & {0.5883} & \bf{0.6310} & 0.6578 & \bf{0.6563} & \bf{0.5923} & \bf{0.5981} & \bf{0.6257} \\
				Quarter & \bf{0.6141} & \bf{0.6059} & {0.5754} & \bf{0.6283} & 0.6537 & \bf{0.6539} & \bf{0.5758} & 0.5906 & \bf{0.6322} \\
				Semester & \bf{0.6222} & \bf{0.6281} & {0.5862} & \bf{0.6435} & 0.6715 & \bf{0.6716} & {0.5910} & 0.6073 & \bf{0.6388} \\
				\midrule
				\multicolumn{10}{c}{\textbf{Matrix factor model}} \\
				\midrule				
				Week  & \bf{0.5690} & \bf{0.5739} & 0.5679 & 0.6009 & 0.6201 & \bf{0.5955} & 0.5660 & \bf{0.5538} & \bf{0.5878} \\
				Month & \bf{0.6138} & 0.6232 & 0.5889 & 0.6352 & \bf{0.6524} & 0.6579 & 0.5947 & 0.5982 & 0.6287 \\
				Quarter & 0.6163 & 0.6138 & 0.5783 & 0.6359 & \bf{0.6477} & 0.6662 & 0.5790 & \bf{0.5899} & 0.6391 \\
				Semester & 0.6225 & 0.6340 & 0.5871 & 0.6499 & \bf{0.6633} & 0.6817 & 0.5922 & \bf{0.6050} & 0.6443 \\
				 \midrule
				\multicolumn{10}{c}{\textbf{Vector factor model}} \\
				\midrule
				Week    & {0.5748} & 0.6222 & 0.5673 & 0.6194 & 0.6223 & 0.6347 & 0.5696 & 0.5671 & 0.6139 \\
				Month   & 0.6210 & 0.6813 & \bf{0.5861} & 0.6567 & {0.6551} & 0.7094 & 0.5961 & 0.6089 & 0.6642 \\
				Quarter & 0.6237 & 0.6872 & \bf{0.5730} & 0.6588 & {0.6501} & 0.7276 & 0.5767 & 0.5922 & 0.6791 \\
				Semester & 0.6291 & 0.7045 & \bf{0.5832} & 0.6699 & {0.6658} & 0.7407 & \bf{0.5907} & 0.6087 & 0.6700 \\
				\midrule
				\multicolumn{10}{c}{\textbf{Functional time series model}} \\
				\midrule
				Week    & 0.6331 & 0.6419 & 0.6388 & 0.6616 & 0.6966 & 0.6397 & 0.6379 & 0.6294 & 0.6497 \\
				Month   & 0.7179 & 0.7411 & 0.7003 & 0.7543 & 0.7959 & 0.7531 & 0.7114 & 0.7234 & 0.7305 \\
				Quarter & 0.8158 & 0.8038 & 0.7652 & 0.8389 & 0.8785 & 0.9364 & 0.7690 & 0.7993 & 0.8142 \\
				Semester & 0.8166 & 0.8025 & 0.7652 & 0.8439 & 0.8852 & 0.8554 & 0.7766 & 0.7972 & 0.8059 \\
				\bottomrule
			\end{tabular}
		\caption{Out-of-sample relative MSE}
		\label{tab:mse_oos_results}
	\end{table}

\newpage
\bibliographystyle{apa}
\bibliography{Bibliography}

\end{document}